\documentclass[fleqn]{2020SCGE}
\begin{document}

\ensubject{subject}

\ArticleType{Article}
\SpecialTopic{SPECIAL TOPIC: }
\Year{2022}
\Month{}
\Vol{}
\No{1}
\DOI{??}
\ArtNo{000000}
\ReceiveDate{September 20, 2022}
\AcceptDate{?? 2022}

\title{Forecast of Cosmological Constraints with Type Ia Supernovae from the Chinese Space Station Telescope}{Forecast of Cosmological Constraints with Type Ia Supernovae from the Chinese Space Station Telescope}

\author[1]{Shi-Yu Li}{}%
\author[2,3]{Yun-Long Li}{}
\author[4,5]{Tianmeng Zhang}{}
\author[6,7,8]{J{\'o}zsef Vink{\'o}}{}
\author[6]{Enik{\H{o}} Reg{\H{o}}s}{}
\author[1,9]{Xiaofeng Wang}{wang\_xf@mail.tsinghua.edu.cn}
\author[9]{\\Gaobo Xi}{}
\author[4,10]{Hu Zhan}{}

\AuthorMark{Wang Xiaofeng}

\AuthorCitation{Li Shi-Yu, Li Yun-Long, Zhang Tianmeng, et al}

\address[1]{Beijing Planetarium, Beijing Academy of Science and Technology, Beijing 100044, PR China}
\address[2]{National Space Science Center, Chinese Academy of Sciences, Beijing 100190, PR China}
\address[3]{National Space Science Data Center of China, Beijing 101407, PR China}
\address[4]{Key Laboratory of Space Astronomy and Technology, National Astronomical Observatories, Chinese Academy of Sciences, 20A Datun Road, Beijing 100101, PR China}
\address[5]{School of Astronomy and Space Science, University of Chinese Academy of Sciences, Beijing 101408, China}
\address[6]{Konkoly Observatory, CSFK, Konkoly Thege M. ut 15-17, Budapest, 1121, Hungary}
\address[7]{ELTE E\"otv\"os Lor\'and University, Institute of Physics, P\'azm\'any P\'eter s\'et\'any 1/A, Budapest, 1117 Hungary}
\address[8]{Institute of Physics, University of Szeged, D\'om t\'er 9, Szeged, 6720, Hungary}
\address[9]{Physics Department and Tsinghua Center for Astrophysics (THCA), Tsinghua University, Beijing 100084, PR China}
\address[10]{ Kavli Institute for Astronomy and Astrophysics, Peking University, Beijing 100871, PR China}


\abstract{The 2-m aperture Chinese Space Station Telescope (CSST), which observes at wavelengths ranging from 255 to 1000 nm, is expected to start science operations in 2024. An ultra-deep field observation program covering approximately 10 square degrees is proposed with supernovae (SNe) and other transients as one of its primary science drivers. This paper presents the simulated detection results of type Ia supernovae (SNe Ia) and explores the impact of new datasets on the determinations of cosmological parameters. The simulated observations are conducted with an exposure time of 150 s and cadences of 10, 20, and 30 days. The survey mode covering a total of 80 observations but with a random cadence in the range of 4 to 14 days is also explored. Our simulation results indicate that the CSST can detect up to $\sim 1800$ SNe Ia at z $<$ 1.3. The simulated SNe Ia are then used to constrain the cosmological parameters. The constraint on $\Omega_m$ can be improved by 37.5\% using the 10-day cadence sample in comparison with the Pantheon sample. A deeper measurement simulation with a 300 s exposure time together with the Pantheon sample improves the current constraints on $\Omega_m$ by 58.3\% and $\omega$ by 47.7\%. Taking future ground-based SNe Ia surveys into consideration, the constraints on $\omega$ can be improved by 59.1\%. The CSST ultra-deep field observation program is expected to discover large amounts of SNe Ia over a broad redshift span and enhance our understanding of the nature of dark energy.}

\keywords{Space-based ultraviolet, optical and infrared telescopes, supernovae; observational cosmology}

\PACS{95.55.Fw, 97.60.Bw, 98.80.-k}

\maketitle


\begin{multicols}{2}
\section{Introduction} \label{sec:intro}

The observations of type Ia supernovae (SNe Ia) have led to 
\Authorfootnote
the discovery of the accelerating expansion of the universe \cite{ref1, ref2}.
This result was later confirmed by other experiments such as the cosmic microwave background (CMB) radiation and baryon acoustic oscillations (BAOs). Although dark energy was invoked to explain cosmic acceleration, its exact nature remains mysterious. SNe Ia are still the most straightforward observational tool to precisely measure the expansion history and hence understand the nature of dark energy \cite{ref3}. 

Over the last two decades, SN surveys have been conducted, such as the ESSENCE Supernova Survey \cite{ref4} that worked on the redshift range of $0.1 \le z \le 0.4$, Supernova Legacy Survey (SNLS) \cite{ref5} that explored the redshift range of $0.3 \le z \le 1.1$, and Dark Energy Survey (DES) \cite{ref6} that worked on the redshift range of  $0.01 < z < 1.2$. The space-based supernova survey Cosmic Assembly Near Infra-Red Deep Extragalactic Legacy Survey and Cluster Lensing And Supernova survey with Hubble \cite{ref7} extended the Hubble diagram beyond $z = 2$. However, current cosmological constraints from SNe Ia still suffer uncertainties due to the limited samples at local and distant universe, especially at z $>$ 1.0. 

The 2-m aperture Chinese Survey Space Telescope (CSST) is expected to be launch-ready by the end of 2023 and start science operation in 2024. The CSST will be in the same low Earth orbit as China's Tiangong Space Station (400 km above the ground), and it will take approximately 90 min to circle the Earth. One of the major scientific goals of the CSST is to constrain the cosmological parameters through multiple probes \cite{ref8}. 

The designed and assumed parameters of the CSST photometric and spectroscopic surveys are described in \cite{ref9, ref10, ref11}. The major survey project of the CSST plans to spend 70\% of the orbital time covering a survey area of 17500 deg$^2$. The number of repeated observations per band within the survey area depends on the number of detectors assigned to that band (2 or 4). Approximately 10\% orbital time is reserved for docking with the space station for maintenance and refueling as needed. Besides the survey camera, the CSST is also equipped with a terahertz receiver, a multichannel imager, an integral field spectrograph, and a cool planet imaging coronagraph. These instruments will share the remaining 20\% orbital time. To detect transient events, such as SNe Ia and superluminous SNe, we propose a 10 square degrees photometric survey program that uses a part of the major survey time. This time-domain survey is complementary to the major survey.

In this paper, we predict the detection of SNe Ia in the proposed 10 deg$^2$ ultra-deep survey sky area and explore the effect of this CSST sample of SNe Ia on cosmological parameters. As the CSST can detect very high-z SNe Ia, the constraints on the cosmological parameters are also investigated with the simulated SN Ia sample. The paper is organized as follows: The strategy of the 10 deg$^2$ ultra-deep survey and the simulations are described in Section \ref{sec:simu}.
The calibration of the simulations is presented in Section \ref{sec:cali}. The cosmology constraints from the simulated survey data of SNe Ia are given in Section \ref{sec:cosmo}. Section \ref{sec:discussion} extends the exposure time to explore the cosmological constraints limit of the CSST time-domain survey. We have also investigated the constraints combined with future ground-based telescope surveys and the current SNe Ia sample. The summary of this work is provided in Section \ref{sec:summary}. 

\section{Simulation of SN Ia detections} \label{sec:simu}

The simulations of SN detections have been performed for some well-known ground-based and space telescopes. For example, \cite{ref12} simulated an SN sample for the ground-based DES and \cite{ref13} conducted a similar analysis for the Wide Field InfraRed Survey Telescope SN survey. Both works adopted the SuperNova ANAlysis (SNANA) package \cite{ref14} in their simulations. The simulation result for the James Webb Space Telescope \cite{ref15} have been recently obtained by SNCosmo \footnote{\url{https://sncosmo.readthedocs.io/en/stable/index.html}} \cite{ref16}, a Python-based tool for supernova cosmology analysis. In this study, we use SNCosmo to simulate the rest-frame spectral energy distributions (SEDs) and generate the light curves of SNe Ia for the CSST.

\subsection{CSST hardware layout and survey strategies}
\label{subsec:strategy}

\begin{figure}[H]
\centering
\includegraphics[scale=1.0,trim=35 15 25 25,clip]{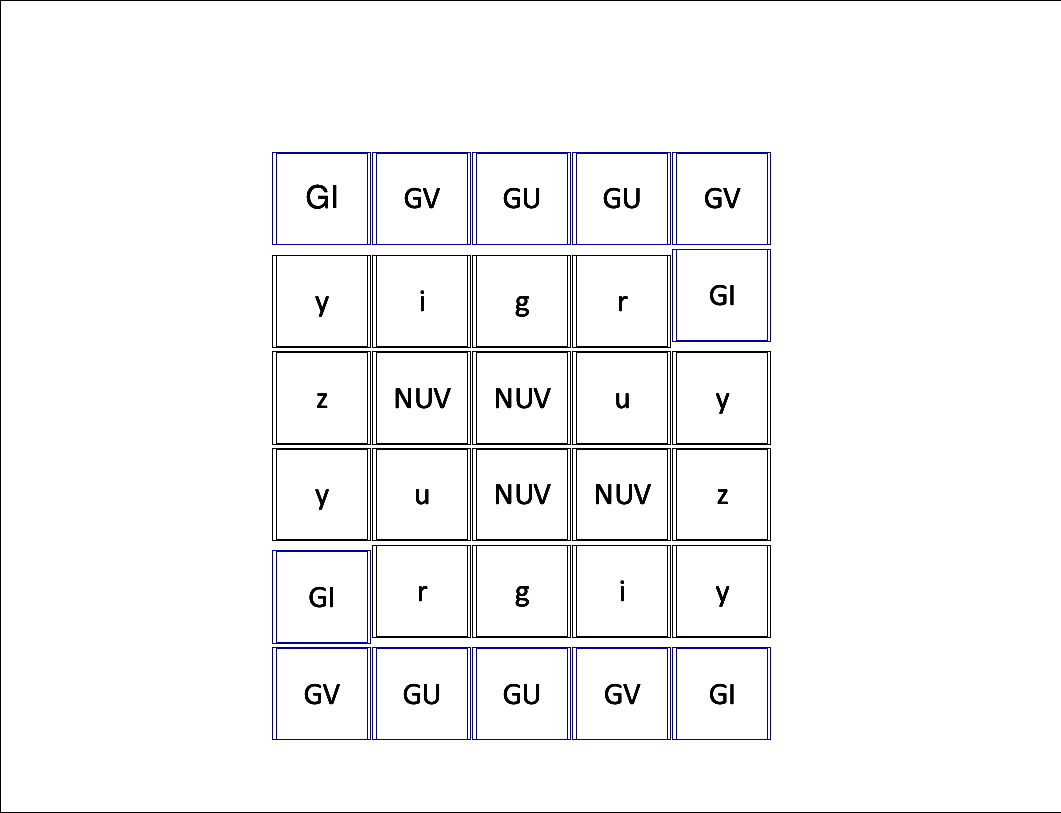}
\caption{Focal plane arrangement of the main-survey camera of CSST. The filters are placed in a $5 \times 6$ array and each filter is equipped with a $\text{9K} \times \text{9K}$ detector. The central field of view (FoV) is approximately 1.1 deg$^2$.
\label{fig:layout}}
\end{figure}

\begin{figure}[H]
\centering
\includegraphics[scale=0.6]{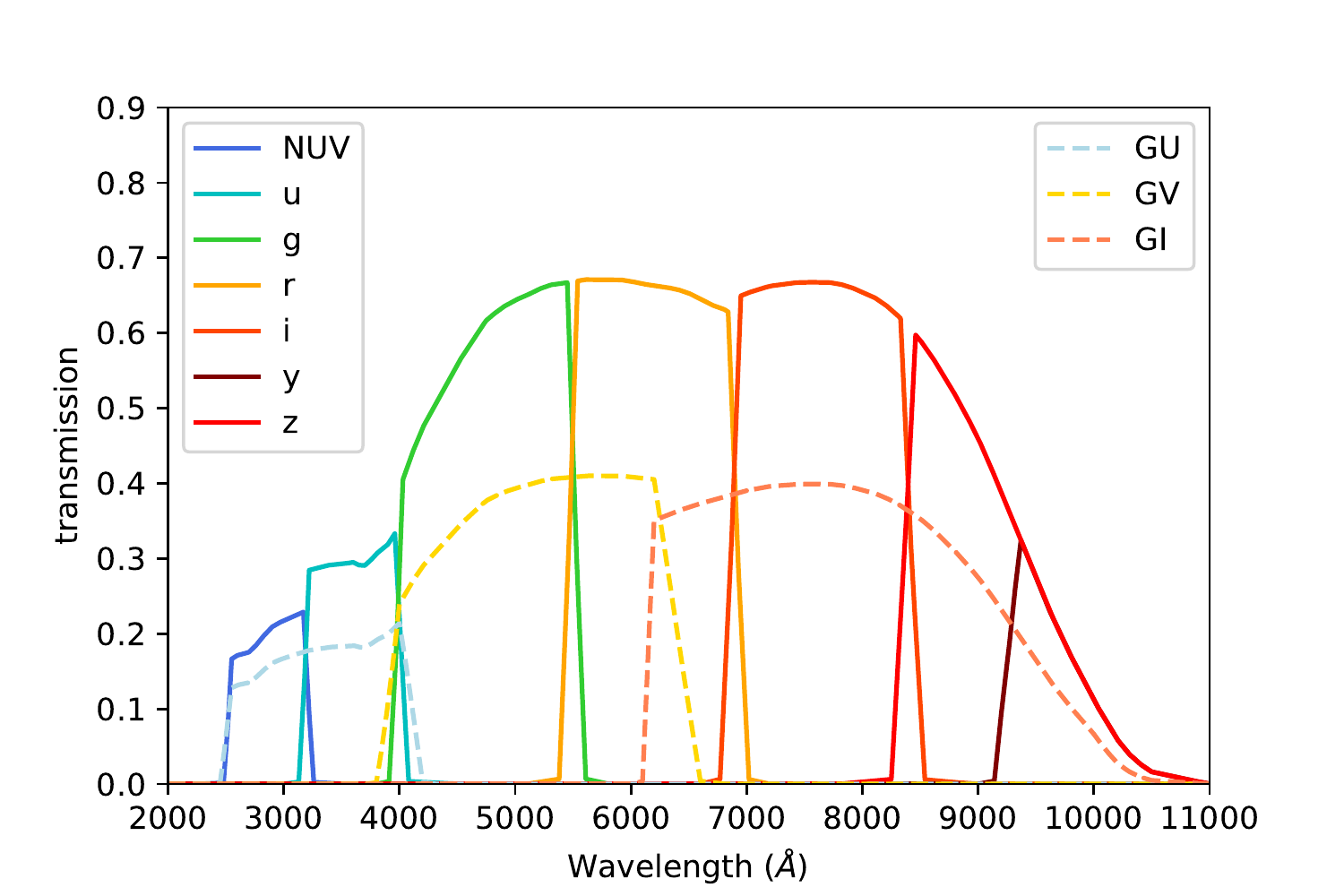}
\caption{Throughput of the CSST photometric system in the NUV, u, g, r, i, z, and y bands (covering the wavelength range from 255 to 1000 nm) and that of the slitless spectroscopic system in the GU, GV, and GI bands.
\label{fig:transmission}}
\end{figure}

The CSST adopts a cook-type off-axis three-mirror anastigmat system. The major instrument of the CSST is a mosaic charge-coupled device camera, which can reach an FoV of approximately 1.1 square degrees with a pixel scale of 0.074 arcsec pixel$^{-1}$. Figure \ref{fig:layout} shows the layout of the focal plane of the main-survey camera. There are 30 9K $\times$ 9K detectors, among which seven imaging filters are used for multi-color photometric observations in the near ultraviolet (NUV), u, g, r, i, z, and y bands mounted on 18 detectors and three gratings are used for slitless spectroscopic observations in the GU, GV, and GI bands mounted on 12 detectors. The transmission curves of each filter are shown in Figure \ref{fig:transmission}. Each filter is equipped with a 9K $\times$ 9K detector. The mean wavelengths and limiting magnitude of the imaging filters are shown in Table \ref{tab:params}. 

\begin{table}[H]
\footnotesize
\begin{threeparttable}\caption{Effective wavelengths and limiting magnitudes for point sources detected at 5$\sigma$ of the imaging filters at an exposure time of 150 s $\times$ 2.}\label{tab:params} 
\doublerulesep 0.1pt \tabcolsep 13pt 
\begin{tabular}{ccc}
\toprule
  	Filter & Effective Wavelength(\AA) & Limiting Magnitude \\\hline
        NUV  & 2884.9 & 25.4 \\
		u & 3685.9 & 25.4  \\
        g & 4708.7 & 26.3 \\
        r & 6093.9 & 26.0 \\
        i & 7556.6 & 25.9 \\
        z & 9059.1 & 25.2 \\
        y & 9825.0 & 24.4 \\
\bottomrule
\end{tabular}\end{threeparttable}
\end{table}

The exposure time for the main survey mission of the CSST is nominally 150 s, whereas the maximum exposure time can reach up to 300 s. For a selected 10 deg$^2$ ultra-deep field, each filter takes a total exposure time of 4.5 h to cover a specific field and completes one epoch observation if the exposure time is 150 s. The exact cadence needs to be determined by considering the schedule of the main survey and the attitude adjustment of the CSST. Ideally, a regular cadence of 10 days is assumed for the following analyses. We then consider a cadence of 20 days and an extreme case with a cadence of 30 days for comparison. Moreover, considering the high priority of the main survey and the possible situation that the CSST will dock with the space station as needed, regular cadence may not be accomplishable. Therefore, a random cadence varying from 4 to 14 days to conduct 80 epochs of observations in 2 years was also simulated.

\subsection{Estimation of the volumetric rate of SNe Ia} \label{subsec:iarate}

The birth rate of SNe Ia as a function of the redshift expected in the 10 deg$^2$ field can be obtained by convolving the cosmic star formation rate (SFR) with the delay time distribution (DTD) derived for SNe Ia, e.g., \cite{ref15}:
\begin{equation}
R_{\text{Ia}}(t) = \upsilon \int_{t_F}^{t}\text{SFR}(t')\text{DTD}(t-t')dt', \label{eq:iarate}
\end{equation}
where $\upsilon$ is a constant coefficient obtained by fitting the observation and $t$ is the cosmic age defined by $t = H_0^{-1}\int_{z}^{\infty} \frac{dz}{(1+z)E(z)}$. $t_F$ represents the formation time of the first star at the redshift $z_F = 10$.

We employ the SFR model from \cite{ref17}:
\begin{equation}
\text{SFR}(z) = K \times \frac{(a+bz)h}{1 + (z/c)^{d}},\label{eq:sfr}
\end{equation}
where $a = 0.017$, $b = 0.13$, $c = 3.3$, $d = 5.3$, and $h =H_0/100 = 0.674$ is given by the Planck satellite \cite{ref18}. The K factor is constrained by the observation data. 

The DTD represents the distribution of the time delay between the birth of the low-mass star and the explosion of the white dwarf in a binary system, which is affected by the supernova explosion model. For example, the DTD given by the single-degeneration scenario tends to be shorter than that by the double-degeneration scenario. We adopt the parameterized model of the DTD from \cite{ref19}:
\begin{equation}
\text{DTD}(t)=\begin{cases}
0 &t<0.04 \,\text{Gyr} \\
7.132\cdot \eta \cdot f_p(1-f_p)^{-1} &0.04 < t < 0.5 \, \text{Gyr} \\ 
\eta \cdot t^{-1} &t > 0.5 \, \text{Gyr}.
\end{cases}
\end{equation}
where $\eta = 2.25$ and $f_p=0.21$. $f_p$ represents the fraction of SNe Ia that exploded within 500 Myr after the formation of their progenitor stars. 
\cite{ref19, ref20} (and references therein) provide observation data to constrain the coefficients in Equation (\ref{eq:iarate}). In the $0 < z < 1.5$ redshift range, the theoretical number of SNe Ia generated with different random seeds is approximately 4300. The redshift distribution of the 4328 simulated SNe Ia for the following analyses is shown in Figure \ref{fig:distr}.

\subsection{Simulations of SN Ia light curves} \label{subsec:Iasimu}

The rest-frame SED of SNe Ia can be modelled using various parameterized models, such as SALT2 \cite{ref21}, MLCS2k2 \cite{ref22}, and SALT3 \cite{ref23}. The SED model we adopted here is a modified SALT2 model \cite{ref24} with the rest-frame wavelength range larger than the fiducial SALT2 model, which is expressed as \cite{ref21}
\begin{equation}
F(t,\lambda) = x_0 \times [M_0(t, \lambda)+x_1M_1(t,\lambda)]\times \text{exp}[-c\times CL(\lambda)],
\label{eq:salt2}
\end{equation} 
where $t$ and $\lambda$ are the rest-frame time and wavelength, respectively; $M_0$ and $M_1$ are the SALT2 vectors that describe the temporal variation of an SN Ia SED; and $x_0$, $x_1$, and $c$ are the free parameters, which could be derived from the fit of SN Ia light curves. $CL(\lambda)$ is the average color correction law, which gives the correction for both the intrinsic color variation and the interstellar extinction. 

In Section \ref{subsec:iarate}, the light curves of SNe Ia in the CSST NUV, u, g, r, i, z, and y bands are simulated for each of the generated SN Ia samples. Five parameters are needed to generate the realistic light curves of SNe Ia, including the time of peak magnitude, redshift distribution, overall scale $x_0$, stretch parameter $x_1$, and color parameter $c$, to generate the simulated SN Ia data. The time of peak magnitude is randomly generated during the two-year observation. The redshift distribution is presented in Section \ref{subsec:iarate}. The SED normalization factor $x_0$ is related to the apparent peak magnitude $m = -2.5\log_{10}[\int F(0,\lambda)\lambda d\lambda]$, which can be derived by sampling the absolute magnitude and redshift. The absolute magnitudes follow a normal distribution of $\mathcal{N}(-19.3,0.212^2)$ \cite{ref25}. $x_1$ and $c$ used in the SALT2 model follow a split normal distribution:
\cite{ref26}:
\begin{equation}
P(x)=\begin{cases}
A \,\text{exp}\left(-\frac{(x-\mu)^2}{2\sigma_1^2}\right), &x<\mu \\
A \,\text{exp}\left(-\frac{(x-\mu)^2}{2\sigma_2^2}\right), &\text{otherwise} \\ 
\end{cases}
\end{equation}
where $A = \sqrt{2/\pi}(\sigma_1+\sigma_2)^{-1}$, and $\mu = 0.938$, $\sigma_1 = 1.551$, $\sigma_2 = 0.269$ for $x_1$, and $\mu = -0.062$, $\sigma_1 = 0.032$, $\sigma_2 = 0.113$ for $c$. The host-galaxy and galactic extinctions are described by the CCM dust model from \cite{ref27}. 

SNe Ia on the Hubble diagram require an accurate determination of the redshift with an error of 0.5\% \cite{ref12}. It is thus crucial to provide follow-up spectroscopic observations either by measuring the spectra of SN Ia or its host galaxy. For host galaxies brighter than magnitude 23, the follow-up observations can be taken by the CSST slitless spectrograph. 

\subsection{Light Curve Fitting} \label{subsec:lc-fit}
Before fitting the light curves, we make selection cuts to the simulated SNe Ia to select high-quality data for cosmological parameter constraints. In the following analyses, the light curves with the maximum signal to noise (SNRMAX) $< 5$ in a given band are considered too weak for detection, and they are thus rejected from the fitting. 
The typical light curve shape of SN Ia rapidly increases in luminosity until it reaches the peak magnitude and then slowly fades away. Therefore, in the fitting process, valid data points (SNRMAX $> 5$) are required before and after the maximum brightness. 
Following the principles of the selection criteria described in \cite{ref12} and \cite{ref13}, we list 
the detection and selection criteria of the simulated SNe Ia for the CSST as follows: 

1. at least one detection with SNRMAX above 10;

2. detection with SNRMAX above 5 in more than one band; 

3. no less than three detections with SNRMAX above 5;

4. at least one detection with SNRMAX above 5 before the B-band peak magnitudes;

5. at least one detection with SNRMAX above 5 after the B-band peak magnitudes.

\begin{figure}[H]
\centering
\includegraphics[scale=0.45]{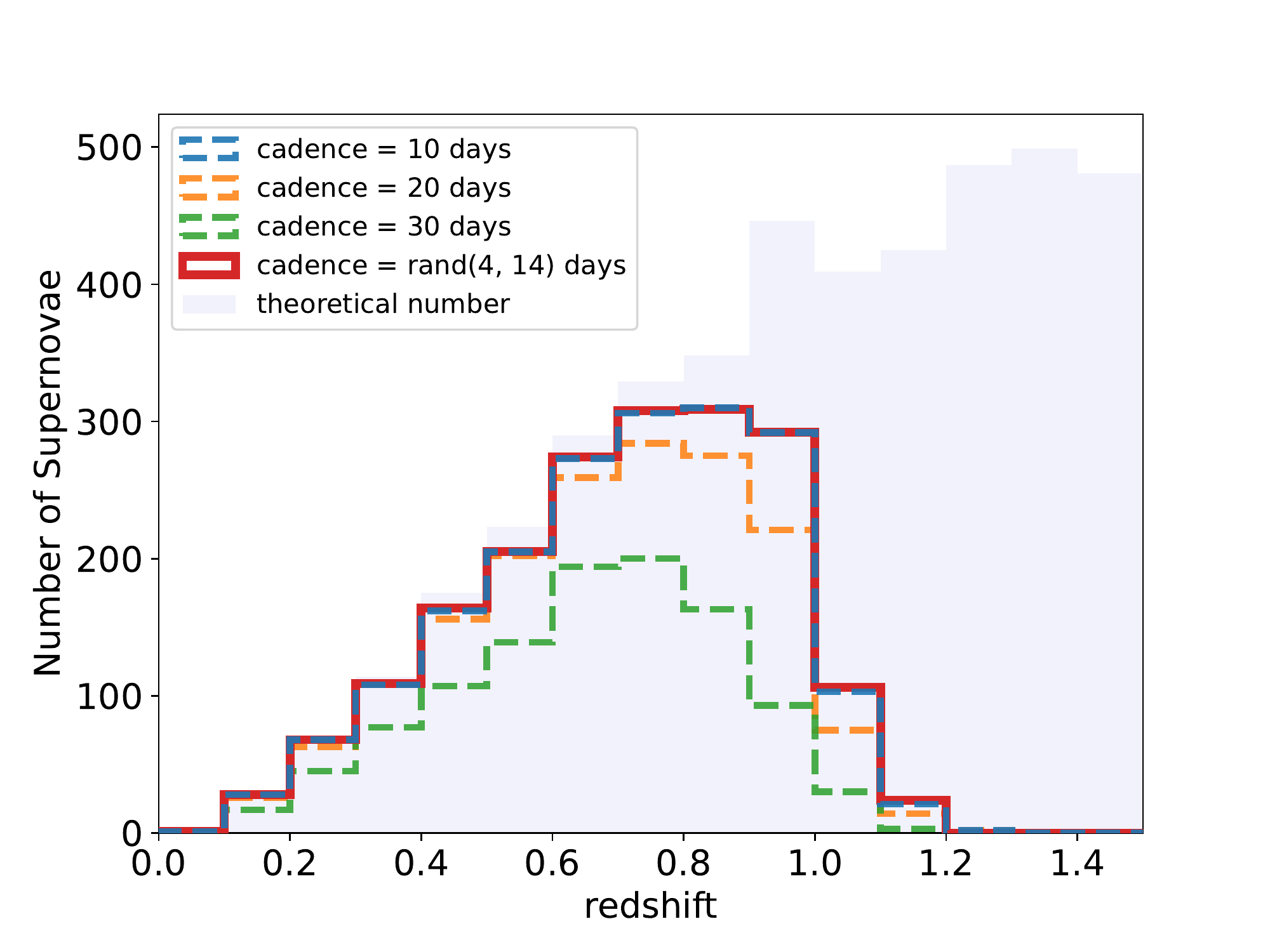}
\caption{Redshift distribution of SNe Ia simulated for the 10 deg$^2$ ultra-deep field of the CSST survey. The lavender-shade region represents the theoretical distribution of SNe Ia obtained in Section \ref{subsec:iarate}, while the blue-dashed, orange-dashed, and green-dashed lines represent the simulated results with a cadence of 10, 20, and 30 days, respectively. The red solid line denotes the survey result with a random cadence ranging from 4 to 14 days.
\label{fig:distr}}
\end{figure}

The total number of SNe Ia passing the above detection and selection criteria is 1879, 1686, 1068 and 1888 for cadences of 10 days, 20 days, and 30 days and the random cadence (ranging from 4 to 14 days), respectively. The redshift distributions of the filtered SNe Ia are shown in Figure \ref{fig:distr}.

Once the light curves pass the detection and selection criteria, they are fitted with the SALT2 model. To ensure that the light curve data are well fitted for the purpose of cosmological analysis, a second filtering criterion is applied in the fitting process, which is described as follows:

1. the fit to the SALT2 model is convergent;

2. the relative error of the apparent magnitude is less than 50\%;

3. the apparent magnitude is brighter than the limiting magnitude of the CSST.

After applying the above additional cuts, the final sample used in our analysis contains 1789, 1619, 1025, and 1802 SNe Ia for cadences of 10 days, 20 days, 30 days, and a random cadence, respectively.

\section{Calibration of the Apparent Magnitude} \label{sec:cali} 

\begin{figure}[H]
\centering
\includegraphics[scale=0.3]{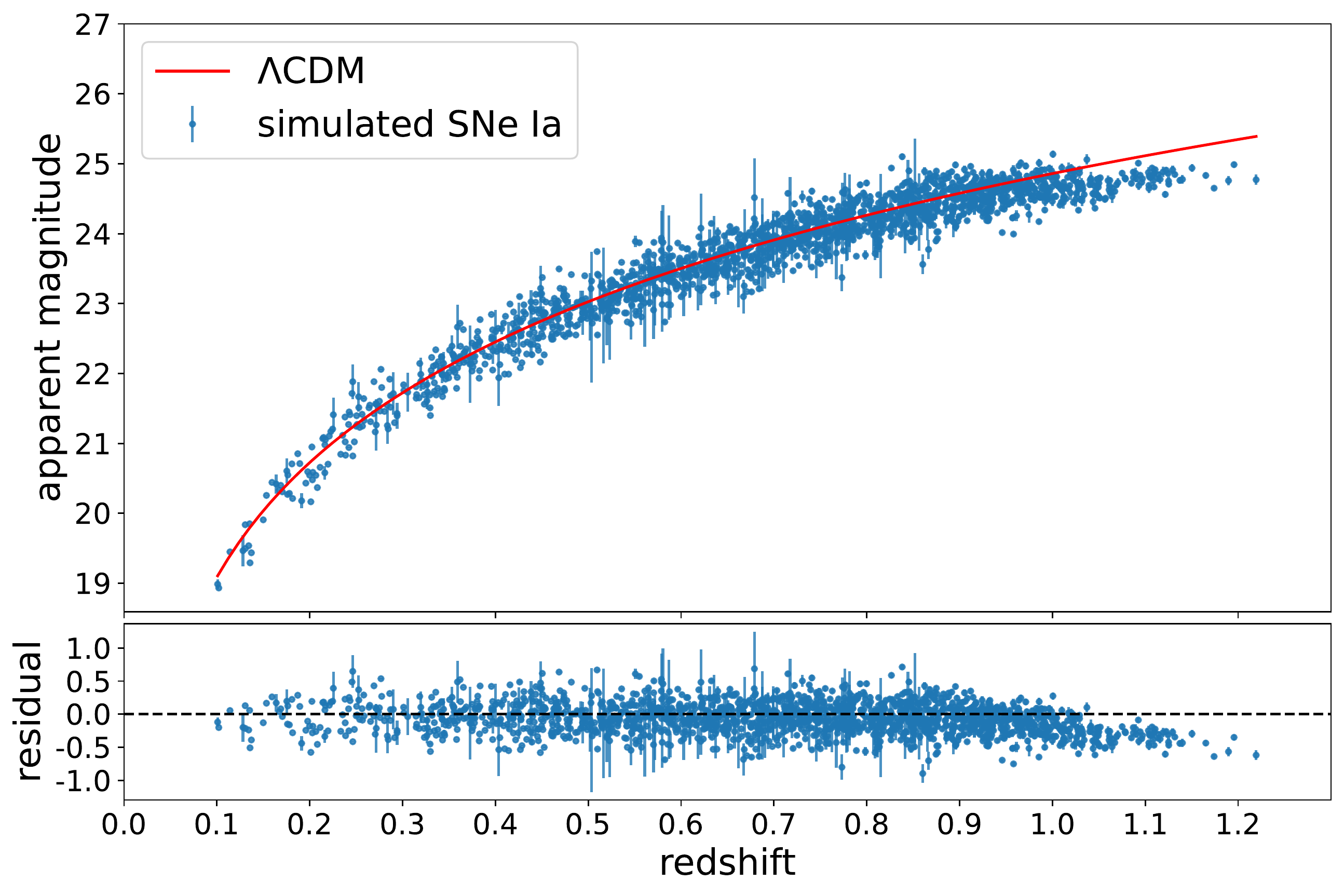}
\quad
\includegraphics[scale=0.3]{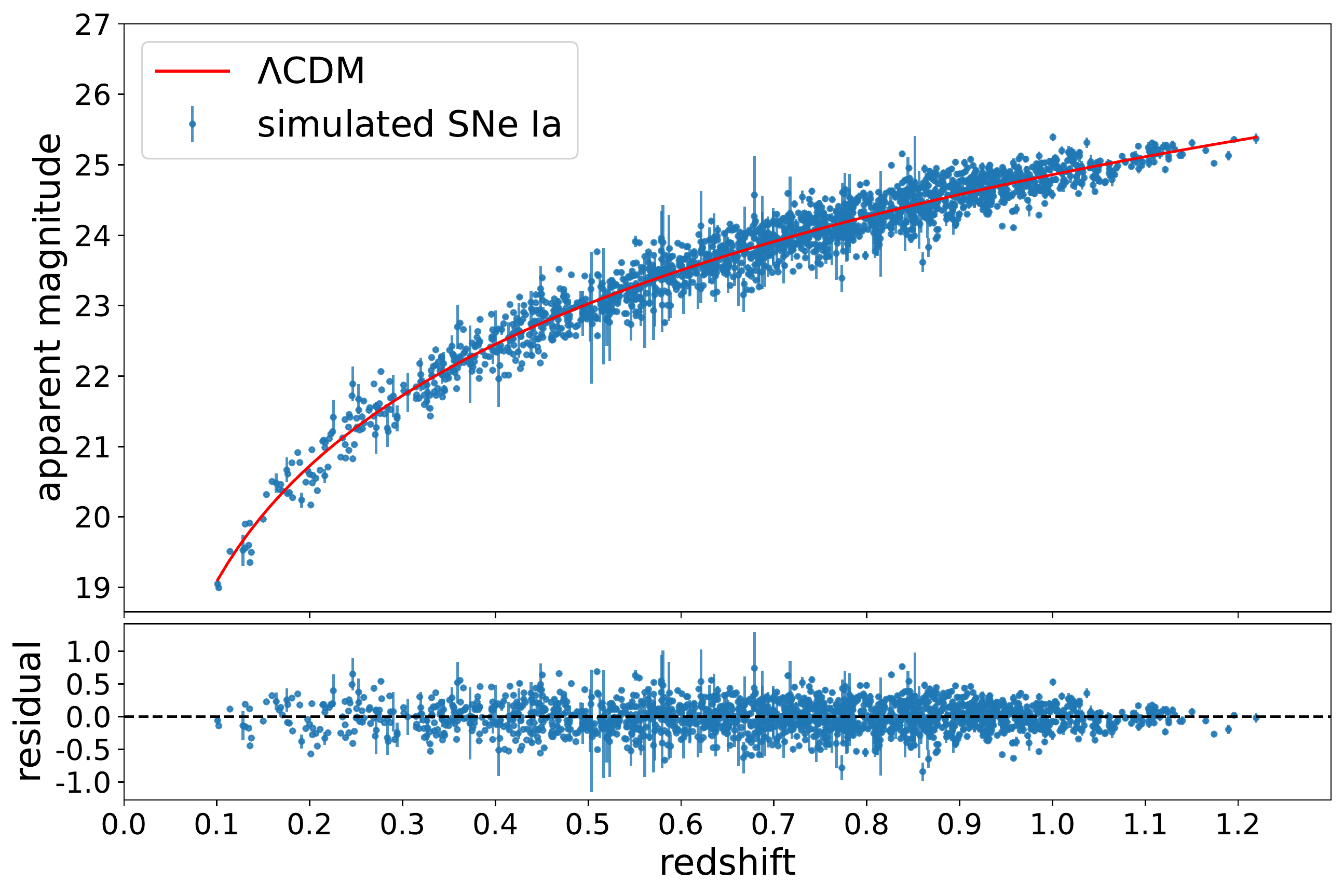}

\caption{Upper panel: Apparent B-band peak magnitude versus redshift derived from the simulation. Lower panel: Apparent B-band calibrated magnitude versus redshift. The sample of SNe Ia is simulated with a regular cadence of 10 days. The red line is the Planck18 $\Lambda$CDM model to guide the eye. Residuals to the Planck18 $\Lambda$CDM model are shown at the bottom of each panel.}
\label{fig:cali}
\end{figure}

The Hubble diagram established for SNe Ia derived from CSST 10 degree$^{2}$ ultra deep field survey with a cadence 
of 10 days is shown in the upper-left panel of Figure \ref{fig:cali}. The scatter relative to the Planck18 $\Lambda$CDM model \cite{ref18} can be attributed to the dispersion of the absolute magnitudes and some artifacts of $H_0$ and $M$ being degenerate of the simulation. Selection effects, such as the Malmquist bias, also make the apparent magnitude at higher redshift tend to be brighter than the theoretical value. We need to preprocess these data before proceeding with the cosmological parameter constraints. This can be obtained by minimizing the $\chi^2$ described in \cite{ref28}:
\begin{equation}
   \chi^2(\alpha, \beta, M_\text{bin}) = \sum_{\text{n}}\frac{\left[ m_{xn} - \mu(z_n) + \alpha x_{1n} - \beta c_n - M_\text{bin}(z_n) \right]^2}{\sigma_n^2+\sigma_{\text{int}}^2}, 
\end{equation}
where $m_{xn}$, $x_{1n}$ and $c_n$ are the best-fit SALT2 parameters of the $n$-th simulated SN Ia and $M_\text{bin}(z)$ is an arbitrary constant in each of the redshift bins. A total of 13 bins with an equal redshift interval of 0.1 are used in this simulation.

The absolute magnitude is defined by 
\begin{equation}
M_\text{bin}(z) = m_x-\mu(z)+\alpha x_1-\beta c,
\end{equation}
where $m_x = -2.5\log_{10}(x_0)$ and $\mu(z)$ is the distance modulus. It is a nearly equivalent formulation of $M_B = m_b - \mu(z)+\alpha x_1-\beta c$, derived from Eq. (\ref{eq:salt2}), where $M_B$ and $m_B$ are the standardized rest-frame B-band absolute magnitude and apparent magnitude, respectively. The difference between $m_x$ and $m_B$ is approximately 10 mag. $\sigma_n^2$ is defined as
\begin{equation}
   \sigma_n^2 = V_{m_{xn}}+\alpha^2 V_{x_{1n}}+\beta^2 V_{c_n} + 2\alpha V_{m_{xn}, x_{1n}} - 2\beta V_{m_{xn}, c_n} - 2\alpha\beta V_{x_{1n},c_n},
\end{equation}
where $V_{n,m_B}$, $V_{n,x_1}$, and $V_{n,c}$ are the variances of $m(x_n)$, $x_{1n}$, and $c_n$, respectively, and $V_{m_{x_n}, x_{1n}}$, $V_{m_{xn}, c_n}$, and $V_{x_{1n},c_n}$ are the off-diagonal elements of the covariance matrix. The intrinsic dispersion we added in this work is $\sigma_{\text{int}}^2=0.045$ mag, which is obtained by iterating the $\chi^2$ converges to the sample size of the CSST-simulated SNe Ia \cite{ref29}. The lower panel of Figure \ref{fig:cali} shows the B-band Hubble diagram with the calibration, and the bottom of each panel shows the residuals to the Planck18 $\Lambda$CDM model.

\section{Constraints on cosmological parameters} \label{sec:cosmo}

The observed apparent magnitudes of SNe Ia and the absolute magnitudes obtained from the SN Ia models allow us to measure their distances. The simulations we obtained in Section \ref{sec:simu} are based on the Planck18 $\Lambda \text{CDM}$ model. To test constraint of our simulated observations on cosmological parameters, we compare the apparent magnitudes to theoretical values adopted from the flat $\omega \text{CDM}$ model. The theoretical apparent magnitude is defined as follows:
\begin{equation}
m_{B,\text{theory}} = \mu + M +K(z)
  = 5\log_{10}\frac{D_L(z)}{\text{Mpc}}+25 + M + K(z),
\label{eq:dl}
\end{equation}
where $K(z)$ is the K-correction for redshift $z$. The K-correction computations are built-in in SNCosmo when converting the rest-frame absolute magnitudes to the observer's frame apparent magnitudes.   

The luminosity distance $D_L$ in Equation (\ref{eq:dl}) under the flat $\omega \text{CDM}$ model is expressed as follows: 
\begin{equation}D_L(z)  = \frac{c_0}{H_0}(1+z)\int_0^z\frac{dz'}{h(z')},
\end{equation}
where $c_0$ is the speed of light and $h(z) = \sqrt{\Omega_m(1+z)^3+(1-\Omega_m)(1+z)^{3(1+w)}}$. The constraints on present matter density $\Omega_m$ and the equation of state of dark energy $\omega$ are found by minimizing the misfit:
\begin{equation}
\begin{aligned}
\chi^2 &= (m_{B,\text{theory}} - m_{B, \text{obs}})^T \mathbf{\Sigma}^{-1} (m_{B, \text{theory}} - m_{B, \text{obs}})^T \\
& = (5\log_{10}Q(z) + \text{const.} + M - m_{B, \text{obs}})^T \mathbf{\Sigma}^{-1}\\ & (5\log_{10}Q(z) + \text{const.} + M - m_{B, \text{obs}})^T, 
\end{aligned}
\end{equation}
where $Q(z) = (1+z)\int_0^z\frac{dz'}{h(z')}$ and $\text{const.} = 5\log_{10}\left(\frac{c}{H_0}\right) + 25$. This work focuses on the constraints of $\Omega_m$ and $\omega$. The absolute magnitude $M$, the Hubble constant $H_0$, and all the constant terms are the nuisance parameters to be marginalized. The modified misfit $\chi^2$ is then given as follows \cite{ref5}:
\begin{equation}
    \chi^2(\Omega_m, \omega) = \boldsymbol{\Delta}^T \left( \mathbf{\Sigma_m}^{-1} - \frac{\mathbf{\Sigma_m}^{-1} \mathbf{F}_1 \mathbf{\Sigma_m}^{T^{-1}}}{\mathbf{1}^T \mathbf{\Sigma_m}^{-1} \mathbf{1}} \right) \boldsymbol{\Delta} + \log{\frac{\mathbf{1}^T \mathbf{\Sigma_m}^{-1} \mathbf{1}}{2 \pi}},
\end{equation}
where $\Sigma_m$ is the covariance matrix of the apparent $B$-band magnitude, $\mathbf{F}_1$ is an $N_{\text{SN}} \times N_{\text{SN}}$ matrix of ones, $\mathbf{1}$ is an $N_{\text{SN}} \times 1$ column vector of ones, and $\boldsymbol{\Delta} =5\log_{10}Q(z)-m_{B,\text{simu}}$. 

Here, we make use of emcee 
\footnote{\url{https://emcee.readthedocs.io/en/stable/}} \cite{ref30} to generate the Markov chain Monte Carlo samples in the posterior space to constrain the cosmological parameters. The likelihood $\text{log}\mathcal{L} \propto -\frac{1}{2} \chi^2(\Omega_m,\omega)$ is analytically marginalized over the nuisance parameters $H_0$ and $M$. For the simulated CSST observations with a cadence of 10 days and exposure time of 150 s, we find the present matter density $\Omega_m=0.35_{-0.06}^{+0.03}$ and the equation of state parameter $\omega = -1.29_{-0.40}^{+0.36}$. The results of the parameter constraints for the other survey strategies with an exposure time of 150 s are consistent with those under a cadence of 10 days but with larger errors. The cosmological constraints are shown in Table \ref{tab:cosmo}.

\section{Discussions} \label{sec:discussion}

In the above analyses, the exposure time of the ultra-deep field of the CSST survey is supposed to be the same as that of the main survey, i.e., 150 s. Here, we perform a similar but deeper exposure simulation with a cadence of 10 days and exposure time of 300 s to explore the limit of the CSST cosmological constraints from SNe Ia. As the maximum volumetric SN Ia rate lies in the redshift range of $1<z<2$ according to various volumetric SN Ia models \cite{ref15}, increasing exposure time will lead to detections of more high-z SNe Ia with a better light curve sampling. Following the method described in Section \ref{sec:simu}, a total of 2660 SNe Ia are within the detection capability of CSST during two years of observations, and 2560 of them pass the fitting criteria. 

\begin{figure}[H]
\centering
\includegraphics[scale=0.32,trim=25 0 0 0,clip]{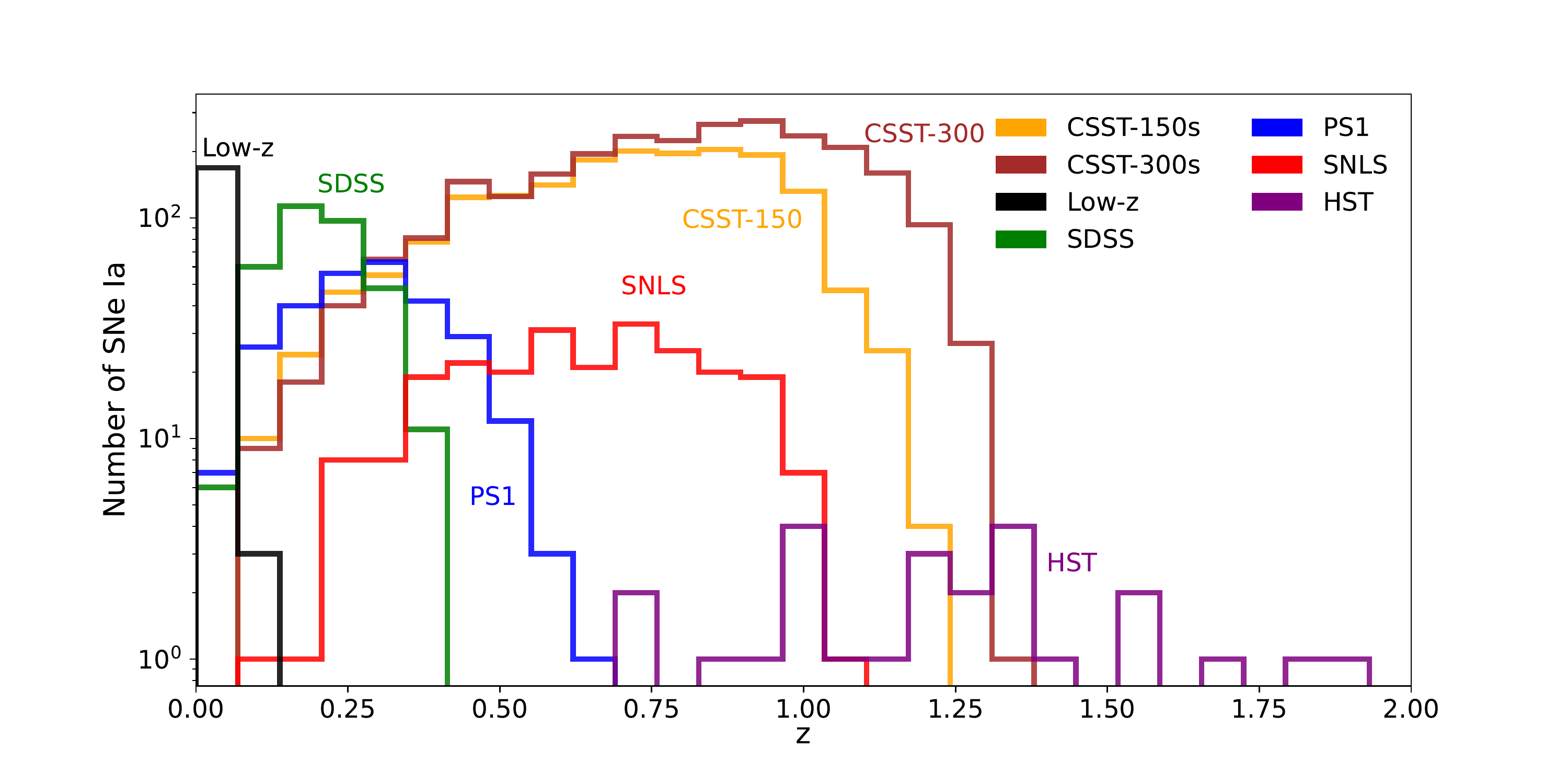}
\caption{Histograms of redshift distribution of detected SNe Ia for the proposed CSST 10 deg$^2$ ultra-deep field survey. The distributions of five subsamples of Pantheon, low-z, SDSS, SNLS, PS1, and HST samples are also plotted. The orange line denotes the distribution predicted for exposure time of 150 s and the brown line denotes that for 300 s exposure. The cadence of the two simulations is 10 days.} 

\label{fig:histz}
\end{figure}

Figure \ref{fig:histz} shows the histogram of the redshift of SNe Ia observed at two different exposure time for the CSST 10 deg$^2$ ultra-deep field survey. The redshift distribution of the Pantheon sample is also shown as a comparison. The lack of low-z SNe Ia in the CSST simulation is attributed to the theoretical volumetric SNe Ia rate. \cite{ref15} showed that the volumetric SNe Ia rates derived from different SFR and DTD models are highly uncertain, which will result in inaccurate quantity estimates. The optimization for the volumetric SNe Ia rate depends on larger SNe Ia samples available for further analysis. Therefore, we added extra low-z ($z<0.1$) samples to give a combined constraint of cosmological parameters.

Ground-based telescopes are expected to discover a large number of local SNe Ia. The Zwicky Transient Facility has discovered over 3000 SNe Ia with a median redshift $\bar{z}= 0.057$ in around 2.5 years \cite{ref31}. From Research to Public outreach telescopes (R2Pub) and wide field survey telescope (WFST) are two new telescopes under construction in China. The R2Pub is a 60 cm-aperture equatorial binocular, and the FoV of each telescope is 18 square degrees. The major scientific goal of the R2Pub is time-domain surveys. The WFST is a 2.5-m wide-field survey telescope with a 3-degree FoV. The Vera C. Rubin Observatory project plans to conduct a 10-year Legacy Survey of Space and Time (LSST) with an 8.4-m Simonyi Survey Telescope. It has a large aperture with a 3-degree FoV and the world's largest digital camera, enabling it to survey the entire sky in only three nights. In the next few years, the R2Pub, WFST, and LSST are expected to detect thousands of nearby SNe Ia. Therefore, we added 3300 low-z ($z<0.1$) samples to the CSST deep measurement simulation dataset. These low redshift samples are generated using the same method as the CSST simulations. 

\begin{figure}[H]
\centering
\includegraphics[scale = 0.42]{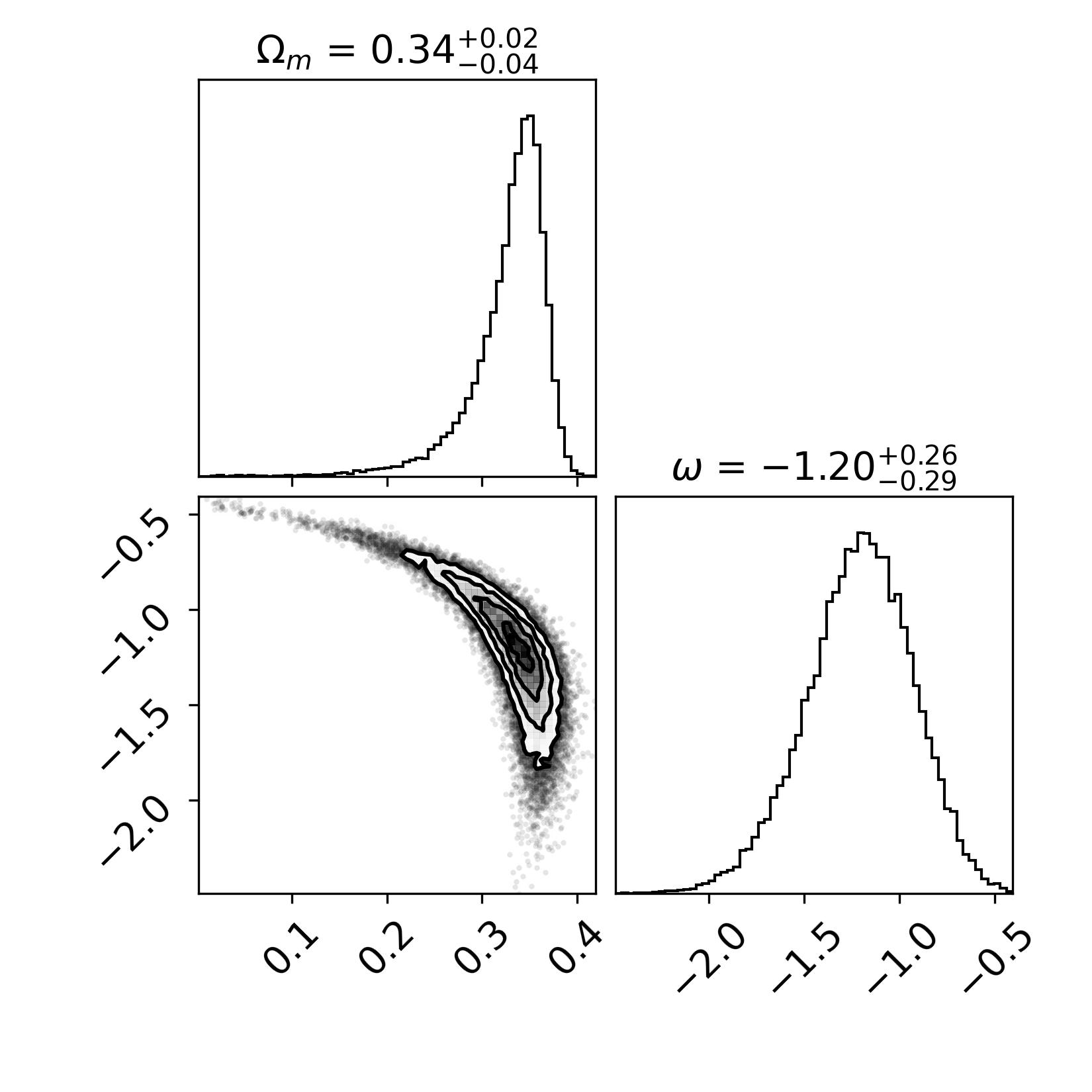}
\quad
\includegraphics[scale = 0.42]{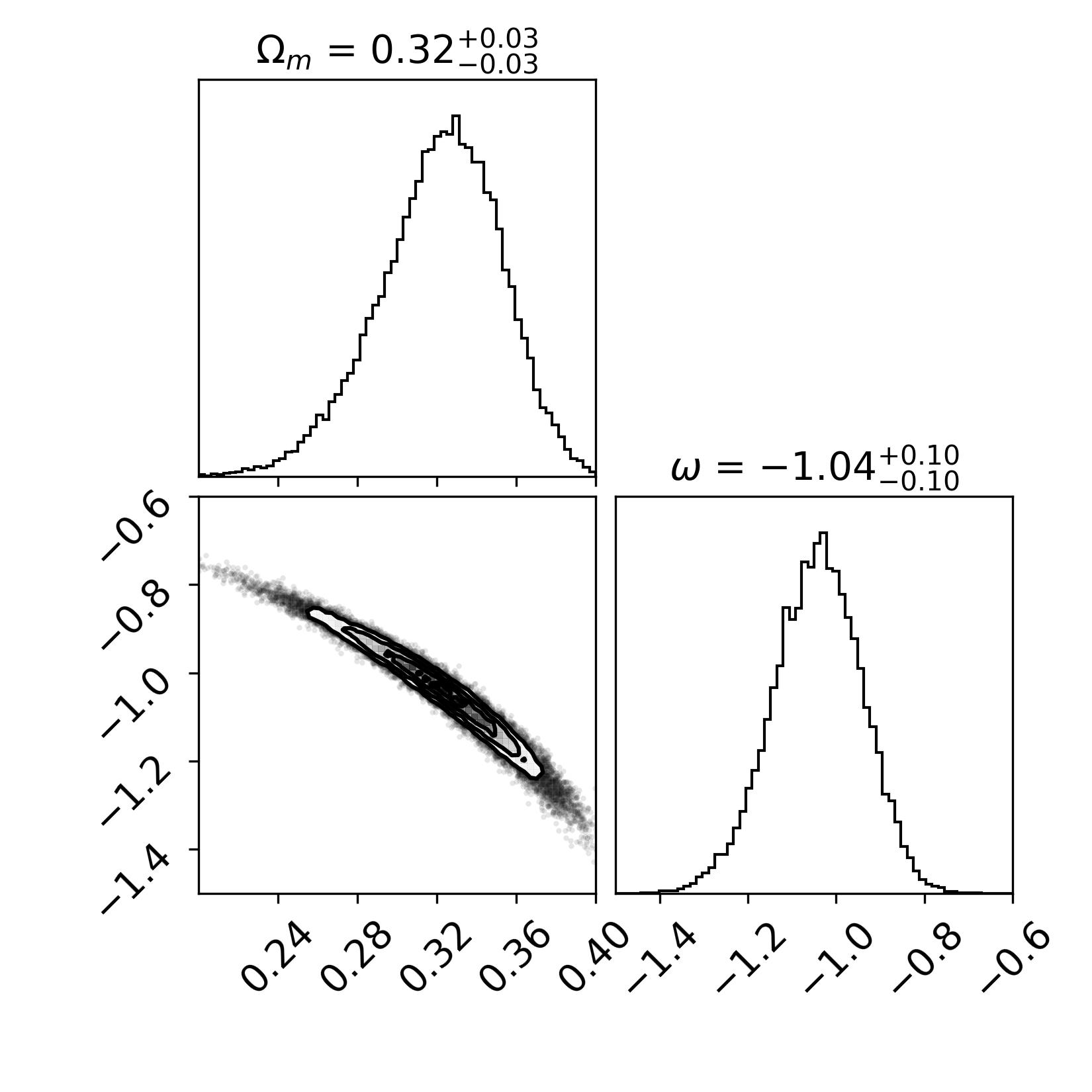}
\quad
\includegraphics[scale = 0.42]{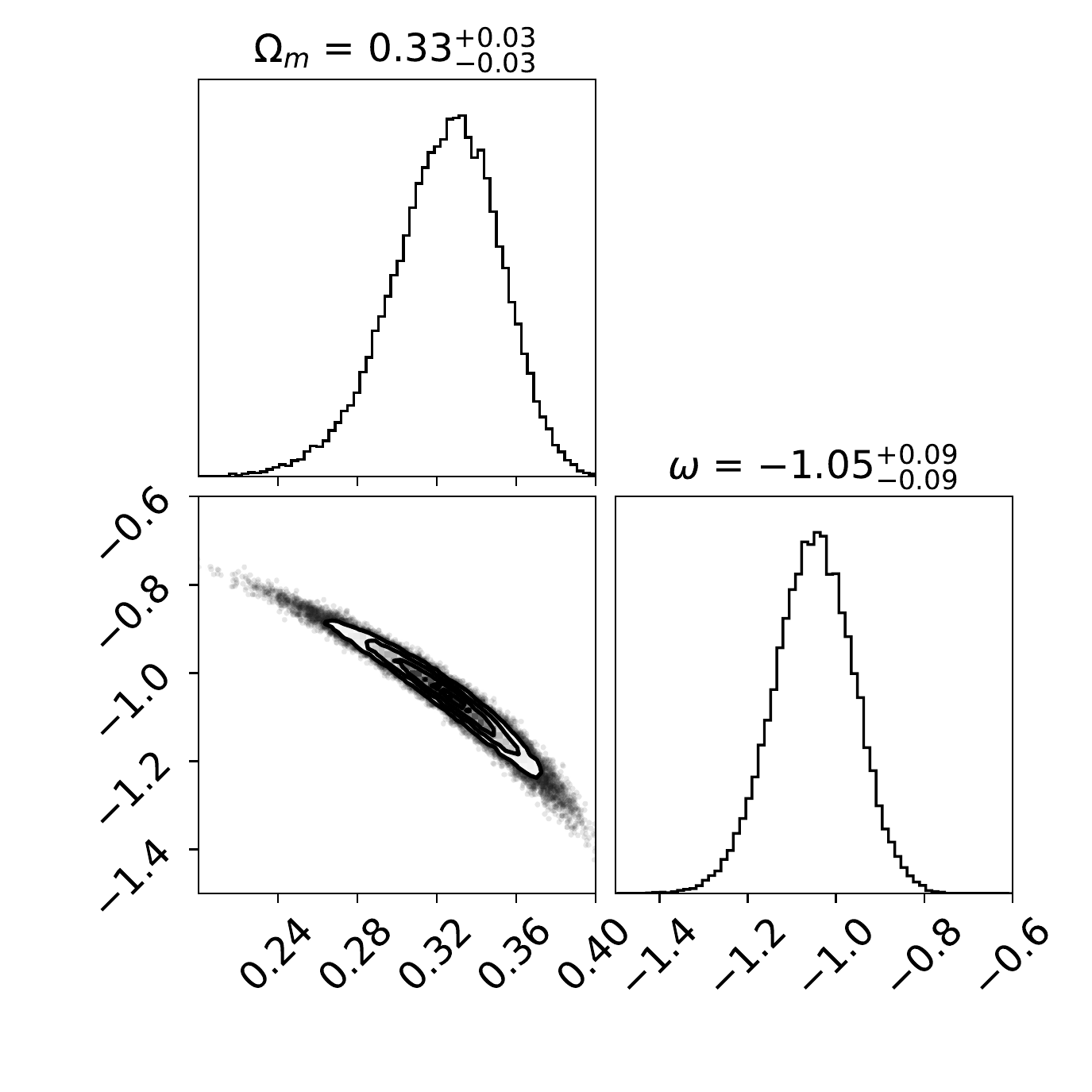}
\caption{Cosmological constraints on present matter density $\Omega_m$ and the equation of state parameter $\omega$. Top panel: simulated cosmological constraints from the CSST sample of SNe Ia. Middle panel: joint constraints of CSST sample with the 3300 simulated low-z SNe Ia sample. Bottom panel:  joint constraints with the simulated low-z sample and Pantheon sample.}
\label{fig:cosmo-mcmc-t300}
\end{figure}

The local SNe Ia data are easier to obtain compared to the high redshift data, and so are the spectra of the SNe themselves or the host galaxies. However, by comparing the retention and removal of the high-redshift SNe Ia in the Pantheon sample, we found that the high-z data ($z>1.3$) play a significant role in the constraints of $\omega$. 

Table \ref{tab:cosmo} shows the results of fitting the $\omega$CMD model to different samples. The deep measurement with an extension of the exposure time improved the constraint on $\Omega_m$ by 33.3\% and $\omega$ by 27.6\% compare to the sample with exposure time of 150 s. Combined with the simulated low-z SNe Ia sample, the constraint on $\omega$ improved by 63.6\% compare to the CSST deep measurement sample. It can be further improved by adding the Pantheon sample, which include high-z data up to $z=2.3$. The corner plots of the cosmological constraints on $\Omega_m$ and $\omega$ of the two configurations are shown in the middle and bottom panels of Figure \ref{fig:cosmo-mcmc-t300}. 

\begin{table*}[t]
\footnotesize
\caption{Cosmological constraints for the $\omega$CDM model. }
\label{tab:cosmo}
\tabcolsep 38pt 
\begin{tabular*}{\textwidth}{lcc}
\toprule
 observation configuration & $\Omega_m$ & $\omega$ \\\hline
 cadence = 10 days, t = 150s & $0.35_{-0.06}^{+0.03}$  & $ -1.29_{-0.40}^{+0.36}$  \\	cadence = 20 days, t = 150s & $0.34_{-0.09}^{+0.04}$  & $ -1.15_{-0.42}^{+0.36}$  \\
 cadence = 30 days, t = 150s & $0.35_{-0.11}^{+0.05}$  & $  -1.15_{-0.53}^{+0.42}$  \\
 cadence = 4-14 days, t = 150s & $0.36_{-0.05}^{+0.03}$  & $ -1.34_{-0.41}^{+0.37}$  \\
 cadence = 10 days, t = 300s & $0.34_{-0.04}^{+0.02}$  & $ -1.20_{-0.29}^{+0.26}$  \\
 cadence = 10 days, t = 300s + low-z sample & $0.32_{-0.03}^{+0.03}$  & $ -1.04_{-0.10}^{+0.10}$  \\
 cadence = 10 days, t = 300s + low-z sample + pantheon sample & $0.33_{-0.03}^{+0.03}$  & $ -1.05_{-0.09}^{+0.09}$  \\
\bottomrule
\end{tabular*}
\end{table*}

As we mentioned in Section \ref{subsec:Iasimu}, it is essential to conduct follow-up spectroscopic observations of the SN itself or its host galaxy. The spectroscopic observations of SN Ia provide accurate measurements of the redshift and precise classification. The most commonly used SNe Ia sample in cosmology contains 1048 spectroscopically confirmed SNe Ia \cite{ref32}. However, the spectroscopic observation of an SN requires observation times on large telescopes within a relatively short time window, so the spectroscopic observation of every SN is not possible. Measuring the host galaxy spectrum is easier to achieve. \cite{ref33} investigated the estimation of cosmological distances according to photometrically classified Pan-STARRS SNe and obtained the cosmology constraints. SNe Ia classified by photometric light curves would be contaminated by core collapse (CC) SNe. Machine-learning algorithms have been used to reduce CC contamination \cite{ref34, ref35}. With the use of these techniques, the CC contamination can be significantly reduced. 

\section{Summary} \label{sec:summary}
In this work, we simulate different survey strategies for the CSST 10 deg$^2$ ultra-deep field survey program to estimate the detections of high-redshift SNe Ia. Based on the theoretical volumetric SNe Ia rate described in Section \ref{subsec:iarate}, up to a total of 1802 SNe Ia are expected to be detected in two years under a survey schedule with a 10-day cadence and 150 s exposure time. A similar number of SNe Ia is expected for a random cadence varying from 4 days to 14 days and a total of 80 revisits in two years. This allows us to be more flexible in scheduling the observations without affecting the main survey project. For an exposure time of 150 s, the redshift range of the detected SNe Ia is from around 0.10 to 1.21. Increasing the exposure time to 300 s extends the detection range to 0.08-1.33. 

The cosmological constraints from the sample collected with a cadence of 20 days are similar to those with a cadence of 10 days and random cadence, although the total observational time spent on the deep fields can be reduced by approximately 50 percent. This result can be considered for future CSST 10 deg$^2$ ultra-deep field survey program. Meanwhile, the CSST SN Ia sample collected with a cadence of 30 days will lead to a significant reduction of useable SNe Ia and hence limited improvements of the cosmological results. 

Cosmological constraint results, particularly the constraints on the dark energy equation of state parameter, will benefit from large amounts of local SNe Ia and high-quality high-z SNe Ia. The proposed CSST 10 deg$^2$ ultra-deep field survey program is expected to enrich SNe Ia sample and explore the nature of the dark energy driving the accelerating expansion of the universe. With SNe Ia alone, the CSST sample, together with the simulated low-z sample and the Pantheon sample, will improve the current Pantheon sample constraints on $\Omega_m$ by 58.3\% and $\omega$ by 59.1\%, respectively. This accuracy can be further improved through a joint analysis of other cosmological probes, such as CMB \cite{ref18}, gravitational-wave \cite{ref36}, BAO\cite{ref37}, and quasars\cite{ref38}.

\Acknowledgements{This work is supported by China Manned Spaced Project (CMS-CSST-2021-A12, CMS-CSST-2021-B01, CMS-CSST-2021-B04), the National Science Foundation of China (NSFC grants 12033003), the Scholar Program of Beijing Academy of Science and Technology (DZ:BS202002), and the Tencent Xplorer Prize. Shi-Yu Li is also supported by Beijing Postdoctoral Research Foundation. Yun-Long Li is supported by the Informatization Plan of Chinese Academy of Sciences, Grant No. CAS-WX2021PY-0101. T.-M. Zhang is supported by the State Key Program of National Natural Science Foundation of China (grant 12233008). JV and ER are supported by the project "Transient Astrophysical Objects" GINOP 2.3.2-15-2016-00033 of the National Research, Development and Innovation Office (NKFIH), Hungary, funded by the European Union. We acknowledge the Space Science Data Fusion Computing Platform of the National Space Science Center for providing computing services. This research made use of Astropy,\footnote{http://www.astropy.org} a community-developed core Python package for Astronomy \cite{ref39, ref40}.}





\end{multicols}
\end{document}